\documentclass[10pt,twoside,twocolumn,english,aps,manuscript,preprint,showpacs,superscriptaddress]{revtex4}
\usepackage[T1]{fontenc}
\usepackage[latin9]{inputenc}
\usepackage{color}
\usepackage{amstext}
\usepackage{graphicx}

\makeatletter

\providecommand{\tabularnewline}{\\}

\@ifundefined{textcolor}{}
{%
 \definecolor{BLACK}{gray}{0}
 \definecolor{WHITE}{gray}{1}
 \definecolor{RED}{rgb}{1,0,0}
 \definecolor{GREEN}{rgb}{0,1,0}
 \definecolor{BLUE}{rgb}{0,0,1}
 \definecolor{CYAN}{cmyk}{1,0,0,0}
 \definecolor{MAGENTA}{cmyk}{0,1,0,0}
 \definecolor{YELLOW}{cmyk}{0,0,1,0}
}

\makeatother

\makeatother

\usepackage{babel}
\begin{document}

\title{Spin liquid behaviour in $J_{eff}=1/2$ triangular lattice Ba$_{3}$IrTi$_{2}$O$_{9}$}

\author{Tusharkanti Dey}

\affiliation{Department of Physics, Indian Institute of Technology Bombay, Powai,
Mumbai 400076, India}

\author{A.V. Mahajan}

\email[Email: ]{mahajan@phy.iitb.ac.in}

\affiliation{Department of Physics, Indian Institute of Technology Bombay, Powai,
Mumbai 400076, India}

\author{P. Khuntia}

\affiliation{Max Planck Institute for Chemical Physics of Solids, 01187 Dresden,
Germany}

\author{M. Baenitz}

\affiliation{Max Planck Institute for Chemical Physics of Solids, 01187 Dresden,
Germany}

\author{B. Koteswararao}

\affiliation{Center for Condensed Matter Sciences, National Taiwan University,
Taipei 10617, Taiwan}

\author{F.C. Chou}

\affiliation{Center for Condensed Matter Sciences, National Taiwan University,
Taipei 10617, Taiwan}
\begin{abstract}
Ba$_{3}$IrTi$_{2}$O$_{9}$ crystallizes in a hexagonal structure
consisting of a layered triangular arrangement of Ir$^{4+}$($J_{eff}=1/2$).
Magnetic susceptibility and heat capacity data show no magnetic ordering
down to $0.35$\,K inspite of a strong magnetic coupling as evidenced
by a large Curie-Weiss temperature $\theta_{\mathrm{\mathrm{CW}}}\sim-130$\,K.
The magnetic heat capacity follows a power law at low temperature.
Our measurements suggest that Ba$_{3}$IrTi$_{2}$O$_{9}$ is a $5d$,
Ir-based ($J_{eff}=1/2$), quantum spin liquid on a $2$D triangular
lattice.
\end{abstract}


\maketitle
\textbf{\textit{Introduction:}} Since Anderson proposed the resonating
valence bond model \cite{Anderson-MRB-8-1973}, researchers have been
searching for experimental realizations of quantum spin liquids (QSL)
\cite{Balents-Nature-464-2010} in geometrically frustrated magnets.
In such materials, incompatibility of local interactions, called frustration,
leads to a strong enhancement of quantum spin fluctuations and effectively
suppresses the long range magnetic ordering. As a result, the material
remains paramagnetic down to very low temperature compared to the
Curie-Weiss (CW) temperature $\theta_{\mathrm{\mathrm{CW}}}$. The
frustration in these materials often arises from some special geometries
like triangular, kagomé, pyrochlore, garnet etc. \cite{Introduction to frustrated magnetism}.

The spin liquid candidates found so far have been mostly $3d$ transition
metal based materials. A few examples are two-dimensional ($2$D)
Kagomé systems SrCr$_{9p}$Ga$_{12-9p}$O$_{19}$ ($S=3/2$) \cite{Ramirez-PRL-84-2000(SCGO)},
and ZnCu$_{3}$(OH)$_{6}$Cl$_{2}$ ($S=1/2$) \cite{Helton-PRL-98-2007},
$S=1$ $2$D triangular lattice antiferromagnet NiGa$_{2}$S$_{4}$
\cite{Nakatsuji-Science-2005}, organic materials like $S=1/2$ triangular
lattice $\kappa$-(ET)$_{2}$Cu$_{2}$(CN)$_{3}$ \cite{Shimizu-PRL-91-2003}
etc. There are very few examples of spin liquid systems with $4d$
or $5d$ spins. Na$_{4}$Ir$_{3}$O$_{8}$ \cite{Okamoto-PRL-99-2007},
a $S=1/2$ spin liquid in a three-dimensional ($3$D) hyperkagome
network, is probably the most notable member of the $5d$ spin liquid
family. 

Recently, Ba$_{3}$CuSb$_{2}$O$_{9}$ ($S=1/2$) with hexagonal space
group P6$_{3}$mc was suggested to be in the QSL ground state \cite{Zhou-PRL-106-2011}.
High pressure hexagonal (P6$_{3}$mc, $6$H-B) and cubic (Fm-$3$m,
$3$C) phases of Ba$_{3}$NiSb$_{2}$O$_{9}$ have also been suggested
to be in the $2$D and $3$D QSL ground state, respectively \cite{Cheng-PRL-107-2011}. 

We have been searching for QSL candidates among hexagonal oxides with
$4d/5d$ elements instead of $3d$ elements. The $5d$ materials are
very different from $3d$ materials and thus interesting because of
a weak onsite Coulomb energy but a strong spin-orbit coupling. For
example, Sr$_{2}$IrO$_{4}$ \cite{Kim-PRL-101-2008} and Ba$_{2}$IrO$_{4}$
\cite{Okabe-PRB-83-2011} are reported to be spin-orbit driven Mott
insulators. The magnetic properties of these systems have been described
based on a $J_{eff}=1/2$ state for the Ir$^{4+}$ ion. Among the
various Ir-based compounds, Ba$_{3}$IrTi$_{2}$O$_{9}$ is rather
interesting since it has a chemical formula similar to the Cu and
Ni-based compounds (discussed in the previous paragraph) and it crystallizes
in a hexagonal structure \cite{Bryne-JSSC-2-1970}. However, detailed
structural parameters have not been reported. Bryne \textit{et al.}
\cite{Bryne-JSSC-2-1970} reported magnetic susceptibility of Ba$_{3}$IrTi$_{2}$O$_{9}$
in the temperature range $77-600$\,K. High antiferromagnetic Weiss
temperature ($\mid\theta_{\mathrm{CW}}\mid>400$\,K) was obtained
by them, suggesting that the magnetic Ir$^{4+}$ ions are strongly
coupled with each other. An obvious question arises, do they order
at lower temperatures? If not then, is it a spin liquid system and
a $5d$ analog of Ba$_{3}$CuSb$_{2}$O$_{9}$? 

Here we report preparation, structural analysis, magnetic susceptibility
and specific heat measurements on Ba$_{3}$IrTi$_{2}$O$_{9}$. It
crystallizes in space group P6$_{3}$mc like Ba$_{3}$CuSb$_{2}$O$_{9}$
and 6H-B phase of Ba$_{3}$NiSb$_{2}$O$_{9}$. A large negative $\theta_{\mathrm{CW}}$
is obtained from CW fitting of susceptibility data but no magnetic
ordering is found from susceptibility and heat capacity measurements
down to $0.35$\,K. Magnetic heat capacity follows a power law at
low temperature. This indicates that the system is highly frustrated
and an example of a $5d$ QSL. We suggest that this is the first candidate
of a $5d$ based quantum spin liquid on a $2$D triangular lattice
with $J_{eff}=1/2$.

\textbf{\textit{Experimental Details:}} Polycrystalline sample of
Ba$_{3}$IrTi$_{2}$O$_{9}$ was prepared by conventional solid state
reaction method using high purity ($99.9\%$) starting materials.

Powder x-ray diffraction (XRD) measurements were performed at room
temperature with Cu $K_{\alpha}$ radiation ($\lambda=1.54182\textrm{\AA}$)
in a PANalytical X\textquoteright{}Pert PRO diffractometer. Magnetization
measurements were performed in a Quantum Design SQUID Vibrating Sample
Magnetometer (SVSM). Heat capacity measurements were carried out in
the temperature range $0.35-295$\,K and field range $0-9$\,T in
a Quantum Design Physical Properties Measurement System (PPMS). High
temperature (upto $800$\,K) susceptibility was measured using a
PPMS VSM.

\textbf{\textit{Results and Discussions:}} XRD measurement was done
to check the phase purity of the sample and to determine crystal parameters,
as the parameters were not mentioned in the earlier report \cite{Bryne-JSSC-2-1970}.
The Ru-analog of Ba$_{3}$IrTi$_{2}$O$_{9}$ i.e., Ba$_{3}$RuTi$_{2}$O$_{9}$
has been mentioned in literature and it crystallizes in the hexagonal
P6$_{3}mc$ space group \cite{Maunders-Acta Crysta-61-2005}. On the
other hand, with a different Ir and Ti ratio, Ba$_{3}$TiIr$_{2}$O$_{9}$
has been suggested to crystallize in the space group P6$_{3}/mmc$
\cite{Sakamoto-JSSC-179-2006}. In both these space groups metal-metal
structural dimers ($2$b sites or $4$f site) are separated by the
$2$a site metal plane. In P6$_{3}mc$, the metal ions within the
dimer are ordered while in P6$_{3}/mmc$ space group the metal ions
within the dimers are not ordered. We tried to refine our XRD data
using these space groups and found P6$_{3}mc$ gives better refinement
with a large site sharing by the Ti$^{4+}$ and Ir$^{4+}$ ions (see
supplemental material \cite{SupplMat}). The lattice constants obtained
from refinement are $a=b=5.7214$$\textrm{\AA}$ and $c=14.0721$$\textrm{\AA}$. 

\begin{figure}
\centering{}\includegraphics[scale=0.75]{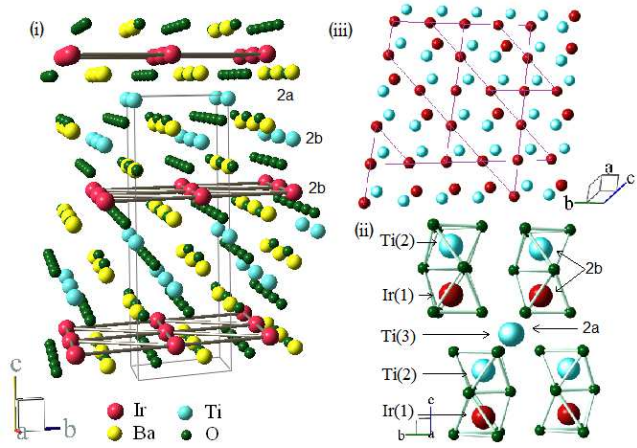}\caption{\label{fig:Structure}(i) Structure of Ba$_{3}$IrTi$_{2}$O$_{9}$
without any site disorder between Ti$^{4+}$ and Ir$^{4+}$. The triangular
arrangement of Ir$^{4+}$ spins in the $ab$ plane is shown (ii) IrTiO$_{9}$
dimers are shown (iii) One possible arrangement of Ti$^{4+}$ and
Ir$^{4+}$ ions in the $ab$ plane is shown when about $1/3$ of Ir$^{4+}$
ions from Ir($1$) site are exchanged with Ti$^{4+}$ ions of the
Ti($2$) site. }
\end{figure}

In the ideal case (i.e., without any site disorder), the Ti($3$)
site is occupied by Ti$^{4+}$ ions and the Ti($2$) and Ir($1$)
sites are occupied by distinct metal ions Ti$^{4+}$ and Ir$^{4+}$
ions, respectively. This is indeed (nearly) the situation in Ba$_{3}$CuSb$_{2}$O$_{9}$
where the Cu site is occupied only by Cu$^{2+}$ (leaving aside a
$5\%$ site disorder), and Sb$^{5+}$ ions are located at Sb($1$)
and Sb($2$) sites. However in our case, we found a $(37\pm10)\%$
site sharing of Ir$^{4+}$ ions with Ti$^{4+}$ ions between Ir($1$)
and Ti($2$) sites and $(7\pm4)\%$ site sharing with Ti$^{4+}$ ions
in Ti($3$) site. This is in fact not unexpected, as their ionic radii
are very similar. Sakamoto \textit{et al.} also found $21\%$ site
sharing between Ti$^{4+}$ and Ir$^{4+}$ in Ba$_{3}$TiIr$_{2}$O$_{9}$
\cite{Sakamoto-JSSC-179-2006} and a similar site disordered situation
is reported in the case of Ba$_{3}$RuTi$_{2}$O$_{9}$ by Radtke
\textit{et al.} \cite{Radtke-PRB-81-2010}. They studied probability
of different Ru$^{4+}$ and Ti$^{4+}$ combinations based on high
resolution electron energy loss spectroscopy and first principles
band structure calculations and concluded that site sharing of ions
in $2$b sites (i.e., Ir($1$) and Ti($2$) sites in our case) is
more probable while site sharing with $2$a sites (i.e., Ir($1$)
and Ti($3$) sites in our case) is less probable. This was suggested
because structural dimers of like ions (Ti-Ti) are energetically unfavourable
due to a strong Ti-Ti repulsion. The same reason is probably valid
in our case and results in a small $7\%$ site sharing between Ir$^{4+}$
ions at Ir($1$) site and Ti$^{4+}$ ions at Ti($3$) site.

In case of perfect ordering among Ti$^{4+}$ and Ir$^{4+}$, these
two ions form face-sharing IrTiO$_{9}$ bioctahedra (shown in Fig.
\ref{fig:Structure}(ii)) and Ir$^{4+}$ spins form an edge-shared
triangular lattice in the $ab$ plane, as shown in Fig. \ref{fig:Structure}.
As a consequence of site disorder, the edge-shared triangular planes
will be depleted. Further, Ir occupying the Ti($2$) sites might also
form a depleted triangular plane. A possible arrangement is shown
in Fig. \ref{fig:Structure}(iii). The blue atoms represent Ti and
the red atoms are Ir. 

\begin{figure}
\centering{}\includegraphics[scale=0.32]{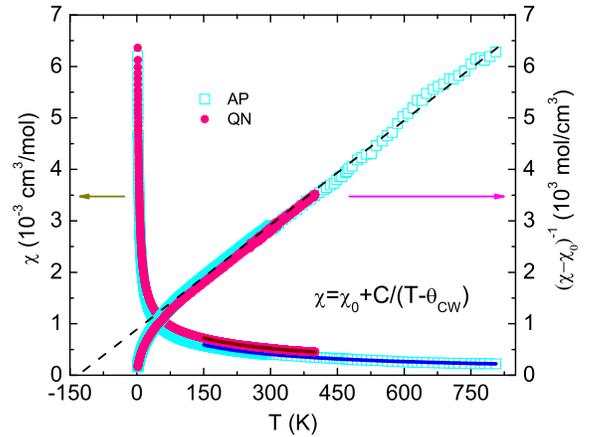}\caption{\label{fig:Chi-T} Left axis: Magnetic susceptibilities of as prepared
(AP) and quenched (QN) Ba$_{3}$IrTi$_{2}$O$_{9}$ samples are shown.
Solid lines denote fitting with CW law in high temperature range $>150$\,K.
Right axis: Inverse susceptibilities (after subtracting $\chi_{0}$)
as function of temperature for both AP and QN. The dashed line is
a linear extrapolation of high temperature data of AP sample.}
\end{figure}

\textcolor{black}{Zero field cooled (ZFC) and field cooled (FC) magnetic
susceptibility was measured with different fields in the temperature
range $2-400$\,K. No magnetic ordering is found down to $2$\,K
but with $100$\,Oe field ZFC-FC splitting is seen below $80$\,K
(shown in Fig. {[}$8${]} in supplemental material \cite{SupplMat}).
However, the splitting is very small (only $11\%$ of total magnetization
at $2$\,K) and supressed when measured even with $500$\,Oe. This
suggests that a small fraction of the spins take part in a glassy
state while the majority of the spins do not. In Na$_{4}$Ir$_{3}$O$_{8}$
also a small ZFC-FC ($<10\%$ of total magnetization) splitting was
observed below $6$\,K which the authors concluded as coming from
a small fraction of the spins \cite{Okamoto-PRL-99-2007}.} Fig. \ref{fig:Chi-T}
shows the temperature ($T$) dependence of dc magnetic susceptibility
of the as-prepared sample (light-blue open squares). Data obtained
\textcolor{black}{with field $5$\,kOe} using a SVSM ($2-300$\,K)
and using a VSM with a high-temperature attachment \textcolor{black}{with
field $50$\,kOe} ($300-800$\,K) have been shown together. Susceptibility
data can be fitted well with the CW formula in the high temperature
($150-800$\,K) region (shown in Fig. \ref{fig:Chi-T}), which yields
\textcolor{black}{temperature independent susceptibility} $\chi_{0}=0.61\times10^{-4}$\,cm$^{3}$/mol,
\textcolor{black}{Curie constant} $C=0.149$\,cm$^{3}$ K/mol and
$\theta_{\mathrm{CW}}=-133$\,K. In many Ir based oxides $\chi_{0}$
is found to be large and varies within a wide range \cite{Sakamoto-JSSC-179-2006,Singh-PRB-82-2010}.
The $C$ value obtained from fitting is much less than that expected
for $S=1/2$ magnetic moments ($0.375$\,cm$^{3}$ K/mol) value.
The large $\theta_{\mathrm{CW}}$ value suggests that there still
are significant correlations in the triangular planes despite the
depletion. The suppression of magnetic moments could be an effect
of the extended nature of the $5d$ orbitals and the strong spin-orbit
coupling expected for $5d$ transition metal oxides. Indeed, in the
magnetically ordered iridates such as Sr$_{2}$IrO$_{4}$ the low
temperature saturation moments have been found to be less than one
tenth of a $\mu_{\mathrm{B}}$ \textcolor{black}{and effective moment
in the paramagnetic region is found to be $\sim0.4\mu_{\mathrm{B}}$}
\cite{Chikara-PRB-80-2009}.

With the aim of investigating as to how the preparation procedure
might affect the site ordering and hence the magnetic properties,
we quenched the as-prepared sample in liquid nitrogen from $1000^{0}$C.
Comparing the normalised x-ray diffraction pattern of both as-prepared
(AP) and quenched (QN) samples, we found that width of all the peaks
and peak height of many peaks are decreased in the QN sample. This
indicates that the crystal symmetry is unchanged but ionic disorder
(and possibly distortions) are less in the QN sample compared to the
AP sample. Refinement of XRD pattern is consistent with a $\sim35\%$
site disorder between Ir$^{4+}$ and Ti$^{4+}$ cations at the $2$b
site but without any site sharing with the $2$a site Ti$^{4+}$ cations.
We also found a marginal increase in the lattice constants with $a=b=5.7216$$\textrm{\AA}$
and $c=14.0768$$\textrm{\AA}$. Susceptibility for the QN sample
\textcolor{black}{measured with field $5$\,kOe} in the temperature
range $2-400$\,K (Fig. \ref{fig:Chi-T}) shows no sign of magnetic
ordering. Data fitted to CW law in the temperature range ($150-400$\,K)
yields $\chi_{0}=1.68\times10^{-4}$\,cm$^{3}$/mol, $C=0.145$\,cm$^{3}$
K/mol and $\theta_{\mathrm{CW}}=-111$\,K. The $\theta_{\mathrm{CW}}$
is somewhat smaller than in the AP sample while $C$ is nearly unchanged.
Inverse susceptibilities (after subtracting the temperature independent
part $\chi_{0}$) of AP and QN samples are linear in temperature and
deviate from linearity below $\sim80$\,K, as shown on the right
axis of Fig. \ref{fig:Chi-T}. 

The large, negative $\theta_{\mathrm{CW}}$ indicates that Ir$^{4+}$
magnetic moments are strongly antiferromagnetically coupled with each
other. Apparently something prevents long range magnetic ordering
to set in even at $0.35$\,K (evident from heat capacity measurement)
which is nearly four hundred times lower than $\theta_{\mathrm{CW}}$.
This suggests that inspite of the depletion of magnetic ions from
the triangular planes, geometrical frustration continues to exist
in the depleted triangular lattice and plays a dominant role in determining
the magnetic properties of this system. Note that a part of the Curie
term could be arising from a few percent of uncorrelated Ir$^{4+}$
spins (and possibly some Ti$^{3+}$ as well) present in the system,
which we call orphan spins (discussed later). 

One should note that in literature $\mid\theta_{\mathrm{CW}}\mid$
and $\mu_{eff}$ reported for Ba$_{3}$IrTi$_{2}$O$_{9}$ are greater
than $400$\,K and $1.73$$\mu_{\mathrm{B}}$, respectively \cite{Bryne-JSSC-2-1970}
which are at variance from our data. To clarify this discrepancy,
we have fitted the published data (Fig. {[}$9${]} in supplemental
material \cite{SupplMat}) with CW law and found $\chi_{0\mathrm{B}}=3.42\times10^{-4}$\,cm$^{3}$/mol,
$C=0.10$\,cm$^{3}$K/mol ($\mu_{eff}=0.89\mu_{B}$) and $\theta_{\mathrm{CW}}=-104$\,K.
Apparently, Bryne \textit{et al.} used $\chi_{0\mathrm{A}}=0.5\times10^{-4}$\,cm$^{3}$/mol
leading them to infer a different $\theta_{\mathrm{CW}}$ and $\mu_{eff}$
(see supplemental material \cite{SupplMat} for details).

\begin{figure}
\centering{}\includegraphics[scale=0.32]{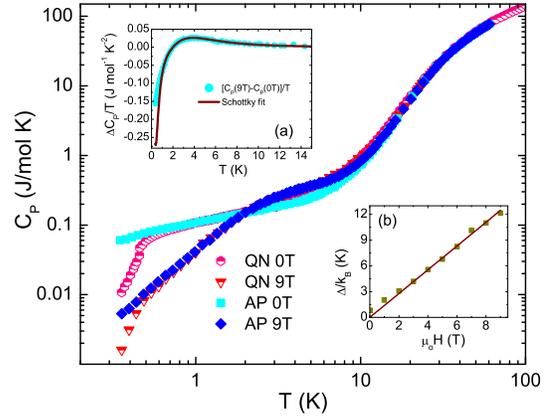}\caption{\label{fig:SpHeat} Heat capacity of AP and QN Ba$_{3}$IrTi$_{2}$O$_{9}$
sample measured in various applied magnetic fields are shown in a
log-log scale. Inset: (a) Solid circles represent {[}$C_{\mathrm{P}}$(9\,T)-$C_{\mathrm{P}}$(0\,T){]}/$T$
of the AP sample and the solid line is the fit (see the text). (b)
$\Delta/k_{B}$ as a function of $\mu_{0}$H from $0$T to $9$T is
shown and the solid line is a fit to Zeeman splitting.}
\end{figure}

Next, in Fig. \ref{fig:SpHeat} we present the heat capacity ($C_{\mathrm{P}}$)
in various fields for the AP and QN samples (data for all fields are
shown in supplemental material \cite{SupplMat}). No anomaly indicative
of long range ordering is found in the measurement range ($0.35-295$\,K).
For both the samples, $C_{\mathrm{P}}$ depends on the applied field
below $\sim20$\,K. This field dependence could be arising from a
Schottky anomaly of orphan spins. We model the heat capacity of Ba$_{3}$IrTi$_{2}$O$_{9}$
as arising out of four contributions. These are namely, the magnetic
contribution of the correlated spins ($C_{\mathrm{M}}$), the lattice
contribution ($C_{\mathrm{lat}}$) and the Schottky anomaly of the
orphan spins ($C_{\mathrm{Sch-orp}}$) and the nuclear Schottky anomaly.

To extract the magnetic part of the heat capacity arising from correlated
magnetic moments, we proceed as follows. The $C_{\mathrm{P}}$ has
contributions from $C_{\mathrm{M}}$, $C_{\mathrm{lat}}$, the Schottky
anomaly ($C_{\mathrm{Sch-orp}}$) from Ir orphan spins and nuclear
Schottky anomaly ($C_{\mathrm{Sch-nuc}}$). $C_{\mathrm{lat}}$ is
field independent while the others might be field dependent. Using
the zero field heat capacity {[}$C_{\mathrm{P}}$($0$\,T){]} and
that measured with `nT' field {[}$C_{\mathrm{P}}$(nT){]}, we obtain
$\Delta C_{\mathrm{P}-\mathrm{Ir}}/T$ = {[}$C_{\mathrm{P}}$(n\,T)-$C_{\mathrm{P}}$(0\,T){]}/$T$.
This is then fitted with $f[C_{\mathrm{Sch}}(\Delta_{1})-C_{\mathrm{Sch}}(\Delta_{2})]/T$,
where $f$ is the percentage of orphan spins present in the sample.
$C_{\mathrm{Sch}}(\Delta_{1}),$ and $C_{\mathrm{Sch}}(\Delta_{2})$
are the Schottky anomalies from $S=1/2$ spins and $\Delta_{1}$ and
$\Delta_{2}$ are the level splittings with applied magnetic fields
$H_{1}$ and $H_{2}$, respectively. Here,

\begin{equation}
C_{\mathrm{Sch}}(\Delta)=R\left(\frac{\Delta}{k_{B}T}\right)^{2}\frac{exp\left(\frac{\Delta}{k_{B}T}\right)}{\left[1+exp\left(\frac{\Delta}{k_{B}T}\right)\right]^{2}}\label{eq:SchottkyEq}
\end{equation}
where $R$ is the universal gas constant and $k_{B}$ is the Boltzman
constant. Inset (a) of Fig. \ref{fig:SpHeat} shows $\Delta C_{\mathrm{P}-\mathrm{Ir}}/T$
obtained for $0$T and $9$T along with the fit described above. The
good \textcolor{black}{fit above $\sim2$}\,\textcolor{black}{K suggests
that }$C_{\mathrm{M}}$\textcolor{black}{{} is not field dependent at
least above $\sim2$}\,\textcolor{black}{K and all the field dependence
is in $C_{\mathrm{Sch-orp}}$. However, below $\sim2$}\,\textcolor{black}{K,
there is deviation of the fit from the data (this is much larger than
the expected nuclear Schottky anomaly) which suggests the }$C_{\mathrm{M}}$\textcolor{black}{{}
might be field dependent there. T}he fraction of orphan spins $f$
is found to be $\sim3\%$. The Schottky splitting ($\Delta/k_{B}$)
obtained from fitting for different fields is plotted as a function
of field in the inset (b) of Fig. \ref{fig:SpHeat}. Similar analysis
has been reported for Ba$_{3}$CuSb$_{2}$O$_{9}$ \cite{Zhou-PRL-106-2011},
ZnCu$_{3}$(OH)$_{6}$Cl$_{2}$ \cite{Vries-PRL-100-2008}, Y$_{2}$BaNiO$_{5}$
\cite{Ramirez-PRL-72-1994(YBNO)} etc. At zero field also we found
a level splitting of $~1.8$\,K which is unexpected but found in
Ba$_{3}$CuSb$_{2}$O$_{9}$ \cite{Zhou-PRL-106-2011} as well. For
$\mu_{0}H\geq2$T, the Schottky splitting gap follows $\Delta=g\mu_{B}H$,
as expected for free spin Schottky anomalies. The `$g$' value for
orphan spins obtained from the linear fit is $2.06$.

\begin{figure}
\centering{}\includegraphics[scale=0.32]{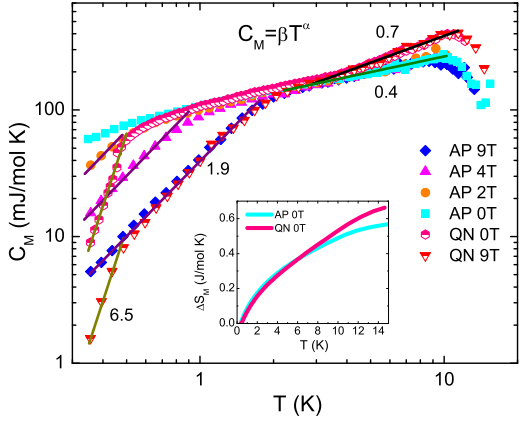}\caption{\label{fig:MagneticHc}Magnetic heat capacity for the AP and the QN
samples are shown. The solid lines are fit to power law with power
indicated in the figure. In the low temperature region, solid lines
with similar color are with same power. Inset: Magnetic entropy change
$\triangle$S$_{M}$ is shown as a function of temperature for $0$T.}
\end{figure}

Using Eq. \ref{eq:SchottkyEq}, the Schottky heat capacity can now
be subtracted from the measured heat capacity of Ba$_{3}$IrTi$_{2}$O$_{9}$.
Next, we would like to extract the lattice heat capacity and for that
we have used Ba$_{3}$ZnSb$_{2}$O$_{9}$ as non-magnetic analog.
\textcolor{black}{Since the Debye frequency is primarily determined
by the lighter atoms (in these cases oxygens), it will not vary much
between these two.} The high-temperature heat capacities of Ba$_{3}$IrTi$_{2}$O$_{9}$
and Ba$_{3}$ZnSb$_{2}$O$_{9}$ differ because of the difference
in their molecular weights and lattice volume. By scaling the heat
capacity of Ba$_{3}$ZnSb$_{2}$O$_{9}$ (obtained from Ref. \cite{Zhou-PRL-106-2011})
by a factor of $\sim0.75$ we find that the heat capacities of Ba$_{3}$ZnSb$_{2}$O$_{9}$
and Ba$_{3}$IrTi$_{2}$O$_{9}$ match in the temperature region $\sim20$\,K-$30$\,K.
The scaled heat capacity of Ba$_{3}$ZnSb$_{2}$O$_{9}$ is then subtracted
from that of Ba$_{3}$IrTi$_{2}$O$_{9}$ in order to obtain the magnetic
heat capacity as shown in Fig. \ref{fig:MagneticHc}. 

\textcolor{black}{The }$C_{\mathrm{M}}$\textcolor{black}{{} for both
AP and QN samples is independent of field from $\sim2.5$-$10$}\,\textcolor{black}{K
and in this range they follow a power law in temperature with power
$\sim0.4$ for the AP sample and $\sim0.7$ for the QN sample. Above
$\sim10$\,K the results can be largely affected by uncertainties
associated with the subtraction process. Notably, }$C_{\mathrm{M}}$\textcolor{black}{{}
for the QN sample is larger than that for the AP sample. Below $\sim2$}\,\textcolor{black}{K,
}$C_{\mathrm{M}}$\textcolor{black}{{} becomes field dependent (for
both samples) but follows a power law with temperature with the same
power for different fields. This power is $1.9$ for the AP sample
and $6.5$ for the QN sample at very low temperature (shown in Fig.
\ref{fig:MagneticHc}). From the heat capacity data of Ba$_{3}$CuSb$_{2}$O$_{9}$
(with space group }P6$_{3}/mmc$\textcolor{black}{) published in Ref.
\cite{Nakatsuji-Science-336-2012}, we have extracted }$C_{\mathrm{M}}$\textcolor{black}{{}
by subtracting a Schottky contribution. Here also we found }$C_{\mathrm{M}}$\textcolor{black}{{}
to be field dependent below $\sim2$}\,\textcolor{black}{K but following
a power law with power $2.1$ for different fields and field independent
in the rage $5-15$}\,\textcolor{black}{K (see }supplemental material\textcolor{black}{{}
\cite{SupplMat}). In many other frustrated systems }$C_{\mathrm{M}}$\textcolor{black}{{}
follows a power law with temperature. The power is $2$ for the $2$D
$S=1$ system NiGa$_{2}$S$_{4}$ \cite{Nakatsuji-Science-2005},
$1$ and $2$ for Ba$_{3}$NiSb$_{2}$O$_{9}$ $6$H-B and $3$C phases
respectively, between $2$ and $3$ for Na$_{4}$Ir$_{3}$O$_{8}$
\cite{Okamoto-PRL-99-2007} and $1$ at low-temperature but $2$ at
higher temperature in $S=1/2$ system Ba$_{3}$CuSb$_{2}$O$_{9}$
(with space group }P6$_{3}mc$\textcolor{black}{) \cite{Zhou-PRL-106-2011}.
A power of $2/3$ was predicted by Motrunich \cite{Motrunich-PRB-72-2005}
for $S=1/2$ triangular lattice organic spin liquid system $\kappa$-(ET)$_{2}$Cu$_{2}$(CN)$_{3}$.
In view of the fact that our Ir-based system is expected to have a
significant spin-orbit coupling, a fresh theoretical effort in this
direction is warranted.}

Magnetic entropy change ($\triangle S_{\mathrm{M}}$) is obtained
by integration of $C_{\mathrm{M}}/T$ with $T$ and is shown as a
function of temperature in the inset of Fig. \ref{fig:MagneticHc}.
The $\triangle$S$_{\mathrm{M}}$ is an order of magnitude lower than
$R$ln$2$ expected for ordered $S=1/2$ systems. In many geometrically
frustrated systems it is observed that the entropy change is lower
than expected value. For example, the entropy change is $30\%$ and
$41\%$ of $R$ln($2S+1$) for Ba$_{3}$CuSb$_{2}$O$_{9}$ and Ba$_{3}$NiSb$_{2}$O$_{9}$
($6$H-B phase) respectively, which are similar in structure with
Ba$_{3}$IrTi$_{2}$O$_{9}$. \textcolor{black}{However, in Ba$_{3}$IrTi$_{2}$O$_{9}$
the magnetic moments are strongly reduced probably due to a strong
spin-orbit coupling. Here $S$ is not a good quantum number and probably
$J_{eff}$ is. So the expected entropy change may not be $R$ln($2S+1$)
i.e $R$ln$2$, but rather a much smaller quantity.} Interestingly,
the heat capacity is different for the QN sample compared to the AP
sample implying the influence of atomic site disorder on the details
of the triangular lattice and hence the ground state.

\textbf{\textit{Conclusions:}} We have presented a potentially new
spin liquid system Ba$_{3}$IrTi$_{2}$O$_{9}$ which is based on
a triangular lattice of Ir$^{4+}$ ions with electrons responsible
for the magnetic properties coming from the $5d$ electronic orbital.
The sample crystallizes in P6$_{3}$mc space group with a large disorder
between Ti$^{4+}$ and Ir$^{4+}$ cations resulting in a site dilution
of nearly $1/3$ of the Ir$^{4+}$ sites of the edge-shared triangular
plane by non-magnetic Ti$^{4+}$. Apparently, magnetic correlations
and frustrations are still maintained with the absence of magnetic
ordering down to $0.35$\,K inspite of a high $\theta_{\mathrm{CW}}$
value ($\sim-130$\,K). Associated with this is a magnetic heat capacity
which, though field dependent, follows a power law with power $1.9$
in the low-temperature range. The QN sample has a different behavior.
This is somewhat like in \textcolor{black}{Ba$_{3}$CuSb$_{2}$O$_{9}$
}where different atomic arrangements (Ref. \cite{Zhou-PRL-106-2011}
and Ref. \cite{Nakatsuji-Science-336-2012}) give rise to different
magnetic heat capacity. As Nakatsuji \textit{et al.} \cite{Nakatsuji-Science-336-2012}
has reported that due to site sharing between Cu$^{2+}$ and Sb$^{5+}$
ions, a distorted honeycomb lattice is formed in Ba$_{3}$CuSb$_{2}$O$_{9}$,
we speculate that a similar situation may occur in Ba$_{3}$IrTi$_{2}$O$_{9}$
yet maintaining a spin liquid ground state. With the demonstration
of the existence of a $J_{eff}=1/2$ state (having a large spin-orbit
coupling) in Sr$_{2}$IrO$_{4}$ \cite{BJ Kim-science-2009}, Ba$_{3}$IrTi$_{2}$O$_{9}$
is possibly an example of a $J_{eff}=1/2$ quantum spin liquid system
and a $5d$ analog of Ba$_{3}$CuSb$_{2}$O$_{9}$. This should open
up a new area pertinent to the search for exotic magnetic behaviour
in $5d$ transition metal based compounds. 

\textbf{\textit{Acknowledgement:}} We thank Department of Science
and Technology, Govt. of India for financial support. FCC acknowledges
the support from National Science Council of Taiwan under project
number NSC-100-2119-M-002-021.

\newpage{}

\textbf{\textcolor{red}{\large Supplemental material for ``Spin liquid
behaviour in $J_{eff}=1/2$ triangular lattice Ba$_{3}$IrTi$_{2}$O$_{9}$''}}\textcolor{red}{\large{} }{\large \par}

..................................................................................

\begin{figure}
\centering{}\includegraphics[scale=0.35]{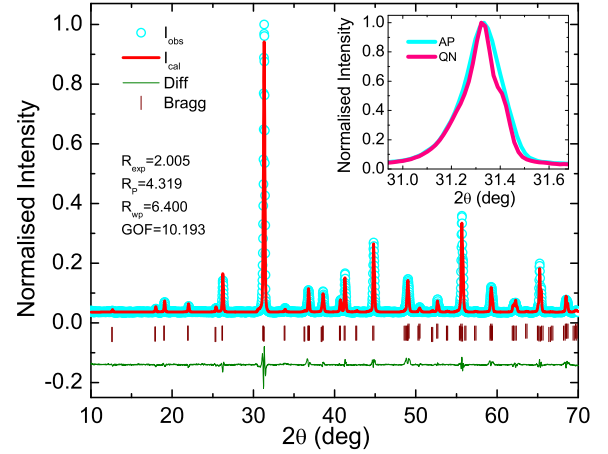}\caption{\label{fig:XRD}Refinement pattern of as-prepared (AP) Ba$_{3}$IrTi$_{2}$O$_{9}$
is shown. Inset: Normalised main peaks for the AP and the quenched
sample (QN) are shown.}
\end{figure}

\begin{figure}
\centering{}\includegraphics[scale=0.35]{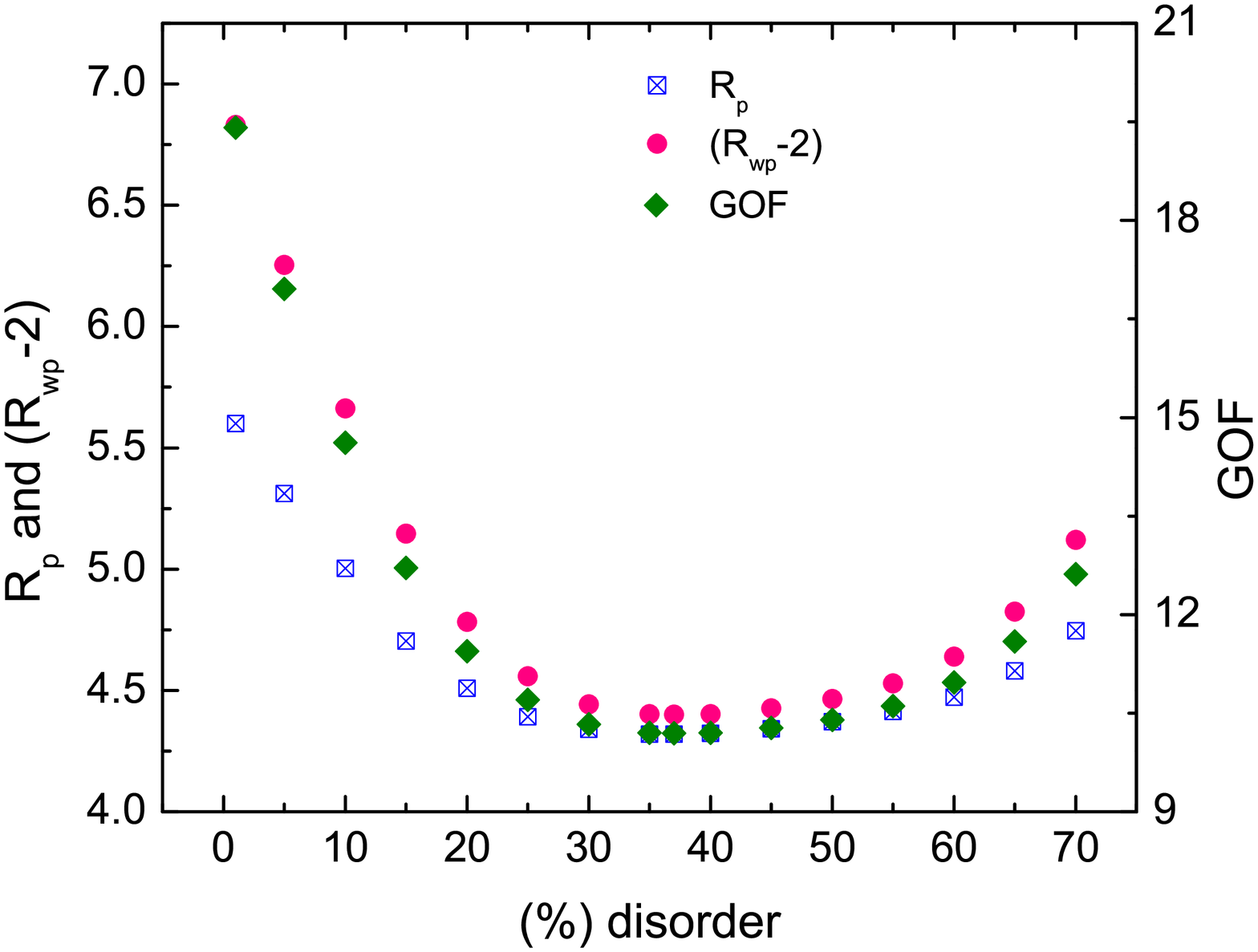}\caption{\label{fig:Ti2sitedisorder} Refinement parameters for the as-prepared
(AP) sample obtained by varying disorder at Ti($2$) site keeping
disorder at Ti($3$ )site unchanged is shown. R$_{p}$ and (R$_{wp}$-$2$)
corresponds to the left axis while goodness of fit (GOF) is plotted
on right axis. R$_{wp}$ has been offset downward by $2$ to show
R$_{p}$ and R$_{wp}$ on the same axis with clarity. }
\end{figure}

\begin{figure}
\centering{}\includegraphics[scale=0.35]{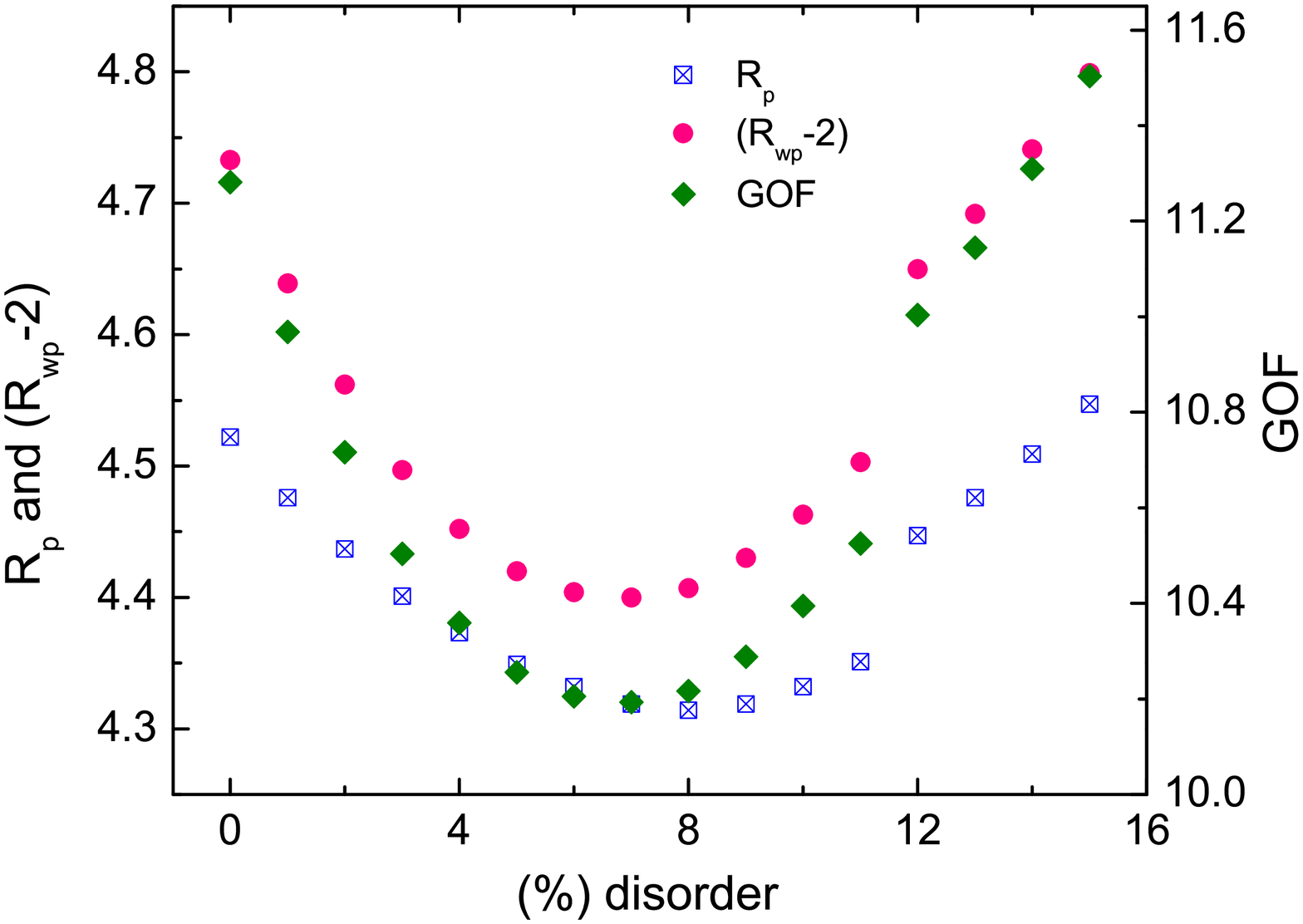}\caption{\label{fig:Ti3sitedisorder}Refinement parameters for the as-prepared
(AP) sample obtained by varying disorder at Ti($3$) site keeping
disorder at Ti($2$) site unchanged is shown. R$_{p}$ and (R$_{wp}$-$2$)
corresponds to the left axis while goodness of fit (GOF) is plotted
on right axis. R$_{wp}$ has been offset downward by $2$ to show
R$_{p}$ and R$_{wp}$ on the same axis with clarity.}
\end{figure}

The XRD refinement of the AP sample is shown in Fig. \ref{fig:XRD}.
The crystal parameters obtained from refinement are given in Table
\ref{tab:XYZ positions}. We have studied the change in the refinement
parameters by varying the site disorder of Ir$^{4+}$ ions with Ti$^{4+}$
ions at Ti($2$) and Ti($3$) sites. The parameters thus obtained
are shown in Fig. \ref{fig:Ti2sitedisorder} and Fig. \ref{fig:Ti3sitedisorder}.
From the figures we can conclude that $(37\pm10)\%$ disorder with
Ti($2$) site and $(7\pm4)\%$ disorder with Ti($3$) site gives best
refinement.

\begin{figure}
\centering{}\includegraphics[scale=0.35]{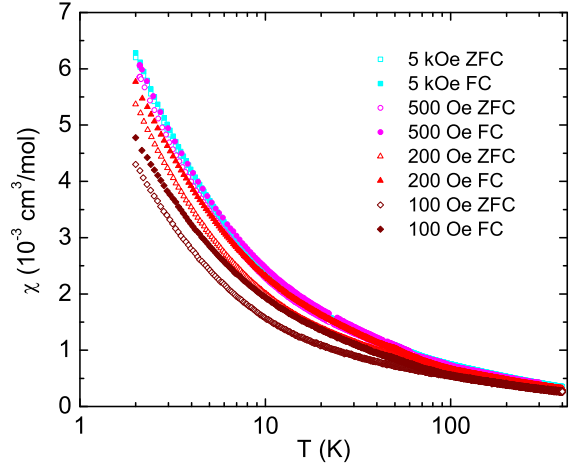}\caption{\label{fig:All Susceptibility} Magnetic susceptibilities of as prepared
(AP) Ba$_{3}$IrTi$_{2}$O$_{9}$ sample measured at different fields
are shown in a semilog scale.}
\end{figure}

ZFC and FC susceptibilities of the AP sample are shown for different
fields in Fig. \ref{fig:All Susceptibility}. 

\begin{figure}
\centering{}\includegraphics[scale=0.35]{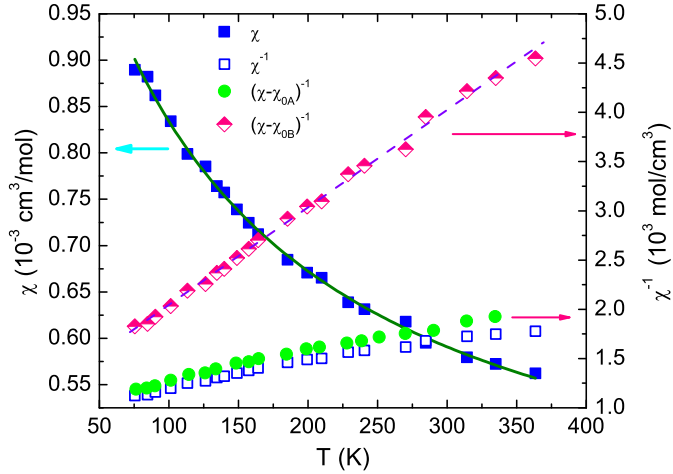}\caption{\label{fig:BryneSusceptibility} Inverse susceptibility data of Ba$_{3}$IrTi$_{2}$O$_{9}$
obtained from Ref. \cite{Bryne-JSSC-2-1970} is shown as blue open
squares. The corresponding susceptibility is shown as blue solid squares
with its fit with Curie Weiss (CW) law (green solid line). Inverse
susceptibility after subtracting $\chi_{0\mathrm{A}}=0.5\times10^{-4}$\,cm$^{3}$/mol
(as in Ref. \cite{Bryne-JSSC-2-1970}) is shown as green solid circles.
The inverse susceptibility after subtracting $\chi_{0\mathrm{B}}=3.42\times10^{-4}$\,cm$^{3}$/mol
(obtained from CW fit done by us) is also shown as pink diamonds.
The dashed line is a guide to eye. Susceptibility is plotted on the
left axis while the inverse susceptibilities correspond to the right
axis.}
\end{figure}

As mentioned in the main paper, the $\mu_{eff}$ and $\mid\theta_{\mathrm{CW}}\mid$
values obtained by Bryne \textit{et al.} \cite{Bryne-JSSC-2-1970}
are much higher than found in our measurements. To find the reason
of this mismatch, we have reanalysed the published data on Ba$_{3}$IrTi$_{2}$O$_{9}$
in Ref. \cite{Bryne-JSSC-2-1970}\textit{.} The blue open squares
in Fig. \ref{fig:BryneSusceptibility} represent the inverse susceptibility
data as published in Ref. \cite{Bryne-JSSC-2-1970}. The corresponding
susceptibility is shown on the left axis as blue solid squares. We
have fitted this susceptibility data with the Curie-Weiss (CW) law
in the range $77-363$\,K. This fitting yields temperature independent
susceptibility $\chi_{0\mathrm{B}}=3.42\times10^{-4}$\,cm$^{3}$/mol,
Curie constant $C=0.10$\,cm$^{3}$K/mol ($\mu_{eff}=0.89\mu_{B}$)
and Curie-Weiss temperature $\theta_{\mathrm{CW}}=-104$\,K which
are somewhat closer to the values obtained from our measurement. Subtracting
this $\chi_{0\mathrm{B}}$, we have plotted ($\chi-\chi_{0\mathrm{B}}$)$^{-1}$
as pink diamonds which is linear in the whole temperature range. Apparently,
Bryne \textit{et al.} did not fit their susceptibility data with CW
law to find $C$ and $\chi_{0}$, rather they have chosen a temperature
independent susceptibility $\chi_{0\mathrm{A}}=0.50\times10^{-4}$\,cm$^{3}$/mol.
We have also shown ($\chi-\chi_{0\mathrm{A}}$)$^{-1}$ data points
as green solid circles which are much different from ($\chi-\chi_{0\mathrm{B}}$)$^{-1}$
data points. The slope corresponding to these green circles has been
used by Bryne \textit{et al.} \cite{Bryne-JSSC-2-1970} to find $\theta_{\mathrm{CW}}$.
Further they have used this $\theta_{\mathrm{CW}}$ and a single (at
$293$K) susceptibility data point ($\chi_{\mathrm{M}}^{'}$) to get
the $\mu_{eff}$ from the formula $\mu_{eff}=2.83\sqrt{\chi_{\mathrm{M}}^{'}(T-\theta)}$.
Hence, fixing $\chi_{0\mathrm{A}}$ (without a fitting procedure)
and calculating $\mu_{eff}$ based on a single susceptibility data
point gives unreliable $\mu_{eff}$ and $\theta_{\mathrm{CW}}$ values
in Ref. \cite{Bryne-JSSC-2-1970}.

\begin{figure}
\centering{}\includegraphics[scale=0.35]{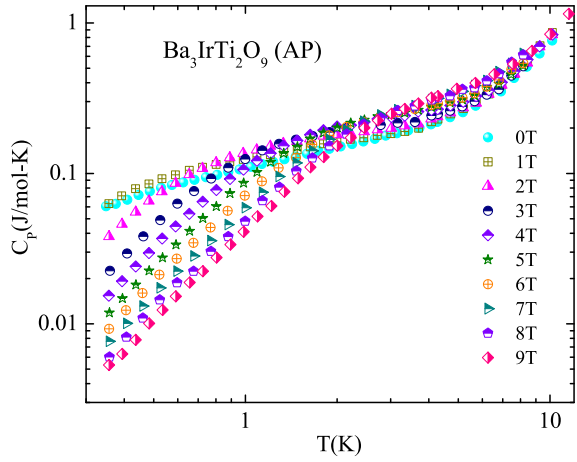}\caption{\label{fig:allCpAP} Heat capacities of as prepared (AP) Ba$_{3}$IrTi$_{2}$O$_{9}$
sample measured in various applied magnetic fields are shown in a
log-log scale.}
\end{figure}

\begin{figure}
\centering{}\includegraphics[scale=0.35]{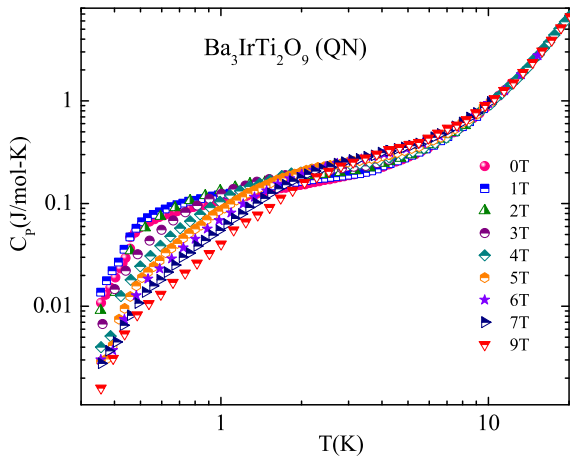}\caption{\label{fig:allCpQN} Heat capacities of quenched (QN) Ba$_{3}$IrTi$_{2}$O$_{9}$
sample measured in various applied magnetic fields are shown in a
log-log scale.}
\end{figure}

\begin{figure}
\centering{}\includegraphics[scale=0.35]{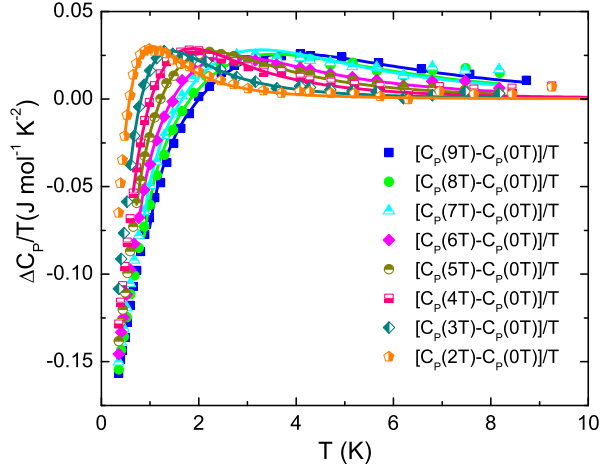}\caption{\label{fig:All Schottky} Scattered symbols represent {[}$C_{\mathrm{P}}$(n\,T)-$C_{\mathrm{P}}$(0\,T){]}/$T$
of the AP sample and the solid line of corresponding color is the
fit (described in the main paper).}
\end{figure}

Heat capacities for different fields for the AP and QN sample are
shown Fig. \ref{fig:allCpAP} and Fig. \ref{fig:allCpQN}, respectively.
Schottky fits for different fields for the AP sample are shown in
Fig. \ref{fig:All Schottky}.

\begin{figure}
\centering{}\includegraphics[scale=0.35]{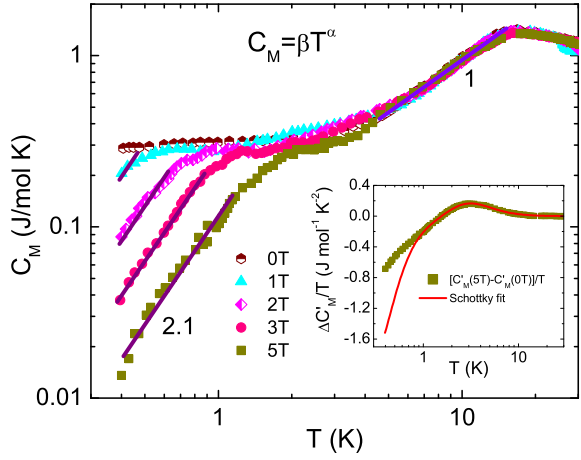}\caption{\label{fig:Nakatsuji-analysis} Magnetic heat capacities ($C_{\mathrm{M}}$)
of Ba$_{3}$CuSb$_{2}$O$_{9}$ (taken from Ref. \cite{Nakatsuji-Science-336-2012})
for different fields after subtracting Schottky contribution are shown.
The solid lines are fits to power law. In the low-temperature region
$C_{\mathrm{M}}$ is field dependent and the power is $2.1$ for different
fields. In the range $5-15$\,K, $C_{\mathrm{M}}$ is independent
of field and linear in temperature. Inset: Solid squares represent
{[}$C'_{\mathrm{M}}$($5$\,T)-$C'_{\mathrm{M}}$($0$\,T){]}/$T$
and the solid line is the fit (see text).}
\end{figure}

In a recent report on Ba$_{3}$CuSb$_{2}$O$_{9}$ (space group P6$_{3}/mmc$),
Nakatsuji \textit{et al. }(Ref. \cite{Nakatsuji-Science-336-2012})
have shown the heat capacity of the system after subtracting the lattice
contribution. We term this as $C'_{\mathrm{M}}$. We have analysed
their data to extract magnetic heat capacity ($C_{\mathrm{M}}$) after
subtracting the Schottky contribution. Inset of Fig. \ref{fig:Nakatsuji-analysis}
shows {[}$C'_{\mathrm{M}}$($5$\,T)-$C'_{\mathrm{M}}$($0$\,T){]}/$T$
and its fit with $f[C_{\mathrm{Sch}}(\Delta_{1})-C_{\mathrm{Sch}}(\Delta_{2})]/T$
(see main paper for details). The fit is good above $\sim1.5$\,K
which means $C_{\mathrm{M}}$ is field independent and the field dependence
of $C'_{\mathrm{M}}$ in that region is totally coming from Schottky
contribution. Below $\sim1.5$\,K, the fit deviates from data points
indicating field dependence of $C_{\mathrm{M}}$. The $\triangle$
obtained from the fit has been used in Eq. $1$ of our paper to get
the Schottky contribution to heat capacity ($f$ was inferred to be
$17\%$). This is subtracted from $C'_{\mathrm{M}}$ to get $C_{\mathrm{M}}$
for Ba$_{3}$CuSb$_{2}$O$_{9}$. For different fields we have extracted
$C_{\mathrm{M}}$ as shown in Fig. \ref{fig:Nakatsuji-analysis}.
$C_{\mathrm{M}}$ is field independent in the range $5-15$\,K and
follows a power law with temperature with power $1$. Below $\sim2$\,K
$C_{\mathrm{M}}$ is field dependent but still follows a power law
with power $2.1$ for different fields. The behavior is very similar
to that found by us in Ba$_{3}$IrTi$_{2}$O$_{9}$.

\begin{table}
\centering{}\caption{\label{tab:XYZ positions}Atomic parameters obtained by refining x-ray
powder diffraction for as-prepared Ba$_{3}$IrTi$_{2}$O$_{9}$ at
room temperature with a space group P6$_{3}$mc. }
\begin{tabular}{|c|c|c|c|c|c|}
\hline 
\multicolumn{1}{|c}{} &  & x & y & z & g\tabularnewline
\hline 
Ba(1) & 2a & 0 & 0 & 0.24199 & 1.00\tabularnewline
\hline 
Ba(2) & 2b & 1/3 & 2/3 & 0.07981 & 1.00\tabularnewline
\hline 
Ba(3) & 2b & 1/3 & 2/3 & 0.39153 & 1.00\tabularnewline
\hline 
Ir(1) & 2b & 1/3 & 2/3 & 0.64464 & 0.56\tabularnewline
\hline 
Ti(1) & 2b & 1/3 & 2/3 & 0.64464 & 0.42\tabularnewline
\hline 
Ti(2) & 2b & 1/3 & 2/3 & 0.83775 & 0.63\tabularnewline
\hline 
Ir(2) & 2b & 1/3 & 2/3 & 0.83775 & 0.37\tabularnewline
\hline 
Ti(3) & 2a & 0 & 0 & 0.48763 & 0.93\tabularnewline
\hline 
Ir(3) & 2a & 0 & 0 & 0.48763 & 0.07\tabularnewline
\hline 
O(1) & 6c & 0.16098 & 0.83898 & 0.57538 & 1.00\tabularnewline
\hline 
O(2) & 6c & 0.48859 & 0.51138 & 0.74989 & 1.00\tabularnewline
\hline 
O(3) & 6c & 0.16098 & 0.83898 & 0.91548 & 1.00\tabularnewline
\hline 
\end{tabular}
\end{table}

\end{document}